\documentclass[sigconf,10pt]{acmart}

\setcopyright{none}

\usepackage[marginal]{footmisc}

\usepackage{balance} 
\usepackage{wrapfig}
\usepackage{comment}
\usepackage{colortbl}
\usepackage[english]{babel}
\usepackage{hyperref}

\usepackage{dblfloatfix} 

\usepackage{amsthm} 
\usepackage{float}
\usepackage{color,soul}
\usepackage{graphics}
\usepackage{hyperref}
\usepackage{lipsum}
\usepackage{tikz}
\usepackage{enumitem}	
\usepackage{listings} 
\usepackage{booktabs}	
\usepackage{lipsum}

\usepackage{capt-of}	
\usepackage{diagbox}
\usepackage{multirow}
\usepackage{todonotes}  
\usepackage{subcaption} 

\definecolor{refkey}{rgb}{249,158,26}
\definecolor{labelkey}{rgb}{0,1,0}
\definecolor{airforceblue}{rgb}{0.36, 0.54, 0.66}
\definecolor{applegreen}{rgb}{0.55, 0.71, 0.0}

\usepackage{mathrsfs}	
\usepackage{amsfonts}

\usepackage[export]{adjustbox} 

\usepackage{algorithm,algorithmic,amsmath} 
\usepackage{boldline}					

\usepackage{multirow}

\definecolor{frenzyorange}{RGB}{249, 158, 26}

\newcommand\paratitle[1]{\textit{\textbf{#1}} \hspace{1mm}}

\renewcommand{\paragraph}[1]{\vskip 3pt\noindent\textbf{#1 }}	 

  {\begin{list}{$\bullet$}%
     {\setlength{\parsep}{0pt}%
      \setlength{\topsep}{0pt}%
      \setlength{\itemsep}{2pt}}}%
  {\end{list}}
\newcommand\Note[1]{\sethlcolor{applegreen} \hl{#1}} 
\newcommand\Noted[1]{} 

\definecolor{darkblue}{rgb}{0.0, 0.0, 0.55}
\definecolor{mygreen}{HTML}{ADFF2F}
\definecolor{mylightgray}{gray}{0.8}

  {\begin{itemize}
	[leftmargin=0cm,
		itemindent=.3cm,
		labelwidth=\itemindent,
		labelsep=0pt,
		parsep=1pt,
		topsep=1pt,
		itemsep=1pt,
		align=left]
  }%
  {\end{itemize}}    

\newenvironment{myenumerate}%
  {\begin{enumerate}
	[leftmargin=.cm,itemindent=.5cm,labelwidth=\itemindent,
		labelsep=0pt,
		parsep=1pt,
		topsep=1pt,
		itemsep=3pt,
		align=left]
  }%
  {\end{enumerate}}    

\newcommand{\tfm}{transformer}

\makeatletter
\def\@copyrightspace{\relax}
\makeatother

\title{Profiling Apple Silicon Performance for ML Training}

\author{Dahua Feng*}
\affiliation{%
	\institution{University of Virginia}
	\city{}
	\state{}
	\country{}
}
\email{wwh8us@virginia.edu}
\author{Zhiming Xu*}
\affiliation{%
	\institution{}
	\city{}
	\state{}
	\country{}
}
\email{zhiming.xu@gmail.com}
\author{Rongxiang Wang}
\affiliation{%
	\institution{University of Virginia}
	\city{}
	\state{}
	\country{}
}
\email{waq9hw@virginia.edu}

\author{Felix Xiaozhu Lin}
\affiliation{%
	\institution{University of Virginia}
	\city{}
	\state{}
	\country{}
}
\email{felixlin@virginia.edu}

\begin{document}


\begin{abstract}

Apple Silicon has attracted much attention for its performance and role in machine learning (ML) training. Unlike NVIDIA GPUs, which have traditionally dominated ML training, Apple Silicon has a significant difference in memory architecture. It uses Unified Memory, which integrates CPU and GPU memory instead of separate CPU memory and GPU VRAM. However, it is difficult to tell whether Unified Memory means more performance benefits.

This paper investigates the performance differences by training several large language model (LLM) workloads end-to-end under different memory scenarios. The results show a significant performance gap between Apple Silicon and NVIDIA GPUs. This paper attributes this gap to system-level factors such as page faults, power consumption, and kernel launch time. In addition, the performance difference of basic linear algebra subprograms (BLAS) on the NVIDIA GPUs and Apple Silicon chips is analyzed to further explain the observed gap.

\end{abstract}
\thispagestyle{empty}   
\maketitle
\pagestyle{plain}
\footnote{* The first author and second author contribute equally to this work.}


\section{Introduction}
\label{sec:intro}


The rapid growth of large language models (LLMs) has expanded machine learning research while introducing significant computational and memory challenges. As LLM sizes increase, their memory requirements often surpass the VRAM limits of high-end GPUs, complicating large-scale training on standard GPU setups. This typically necessitates complex scheduling algorithms, reducing training efficiency. Consequently, access to sufficient training resources has become a major hurdle, particularly for independent researchers and small institutions.


Apple Silicon marks a shift with its M1 and M2 chips, unifying CPU, GPU, and Neural Engine in a unified memory pool of up to 128GB. This architecture addresses VRAM-limited GPU constraints, offering a portable, cost-effective alternative to traditional GPU workstations, and democratizing ML training and research. Besides, software environment support for Apple Silicon is also constantly improving. Pytorch has stable support for the MPS backend since version 1.12~\cite{pytorchPreviousPyTorch}, which provides great support for using Apple Silicon for machine learning tasks. Later, the introduction of the MLX framework~\cite{mlx2023} further optimized the performance of Apple Silicon in machine learning.
However, despite these promising advantages, Apple Silicon devices still face scrutiny for computing power, especially when training LLMs, which require high throughput and large amounts of memory bandwidth. It must be admitted that the performance of Apple Silicon devices during training is poor compared to NVIDIA GPUs, but the reasons for the performance gap are not clear, which has brought resistance to the popularity of Apple Silicon in LLM training.

This paper focuses on the scenario under the single-chip LLM training and fine-tuning, explores the potential and limitations of Apple Silicon, and studies its architectural advantages, memory management capabilities, and performance trade-offs. Ultimately, our goal is to evaluate whether Apple devices like MacBooks can provide a viable and accessible solution for ML training, thereby bridging the gap between professional hardware and consumer-level options. We mainly answer the following questions in this paper: (1) What are the specific advantages and disadvantages of Apple Silicon and NVIDIA GPUs in LLM training? (2) What may be the reasons for these advantages and disadvantages of Apple Silicon? (3) How can Apple Silicon be used for LLM training to achieve more satisfactory performance?
\section{Background}
\label{sec:bkgnd}


\subsection{Hardware Characteristics}
\label{sec:motiv:primer}
%

\paragraph{Apple Silicon.} 
Apple Silicon first launched the M1 chip in 2020~\cite{applem1}, followed by subsequent models such as the M1 Pro, M1 Max, M1 Ultra, and M2 series, setting a new standard for performance and efficiency in consumer devices. Apple Silicon is built on a System-on-Chip (SoC) architecture, integrating multiple components, including the CPU, GPU, and Apple's Neural Engine, into a single unified memory architecture, which is a notable feature of Apple Silicon. This design not only simplifies data processing but also supports shared memory between processing units, significantly reducing latency and energy consumption compared to traditional multi-chip setups.

The unified memory architecture in Apple Silicon is particularly exciting under the context of LLM training. Unlike discrete GPUs that rely on limited capacity dedicated VRAM, Apple Silicon's unified memory provides a more flexible and richer memory pool. Devices like MacBook Pro can support up to 128GB of unified memory, and Mac Studio can support up to 192GB of unified memory, which will be very beneficial for training tasks that require a lot of memory to handle large-scale LLMs.
However, the efficiency and memory pooling advantages of the unified memory can be offset by lower raw compute throughput compared to dedicated GPU workstations, which limits Apple Silicon's competitiveness in large-scale or production-level machine learning workflows. This trade-off sets the stage for a deeper exploration of Apple Silicon's capabilities and limitations, evaluating its viability as a viable platform for accessible machine learning training, especially in the context of LLM and other resource-demanding models.



\subsection{Software Support}

\begin{figure}[h]
	\includegraphics[width=0.4\textwidth]{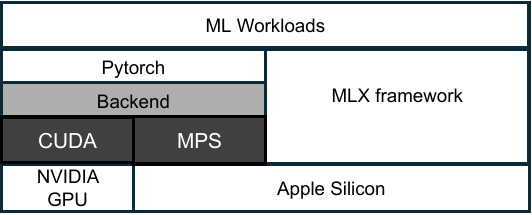}
    
	\caption{The software/hardware stack for ML training, showing for both Apple GPU and NVIDIA GPU.}
	\label{fig:bkg:softwaresupport}
\end{figure}
\paragraph{Pytorch.} With the release of Apple Silicon, Apple introduced the Metal Performance Shaders~\cite{mps} framework, i.e., MPS, a high-performance, low-level API optimized for GPU-accelerated computing on macOS devices. 
The MPS backend enables PyTorch to take advantage of Apple's GPU architecture for efficient memory handling and computational acceleration, making it possible to perform complex machine learning inference or training tasks on devices that were previously limited to CPU processing on macOS. Unlike traditional GPU architectures designed around independent VRAM, Apple Silicon's unified memory architecture provides a shared memory pool accessible to CPU and GPU cores, which may have advantages when managing large LLM models that require a lot of memory capacity. The support for the MPS backend is still being improved. Currently, Pytorch still lacks support for some training-related optimizations, which are already very complete on the support for CUDA devices.

\paragraph{MLX.} MLX is an array framework for efficient machine learning on Apple Silicon supporting Apple machine learning research~\cite{mlx2023}. Specifically, MLX can perform common machine learning operations such as matrix multiplication, convolution, and data transformations faster, which are critical for training and inference in deep learning models. In the software stack, MLX is located in the same position as Pytorch, which means MLX is potentially an alternative for Pytorch in machine learning on the Apple Silicon.
The most significant feature of MLX is its unified memory model, which is also a feature that distinguishes it from the PyTorch with MPS above. Arrays in MLX exist in shared memory, allowing MLX arrays to be operated across supported device types without the need for data transfer. We take the MLX in the scope, with the hope that MLX can have a difference on machine learning using Apple Silicon and we can have a better understanding of the influence of software support on machine learning training performance. 
\section{Method}
\label{sec:method}

\subsection{Choices of hardware}
We use several servers and devices to conduct our experiments. We mainly focus on devices with comparable costs that can be afforded by regular research groups and institutions. For Nvidia devices, we conduct experiments on servers with four distinct models of middle- to high-end consumer and professional GPUs. For Apple Silicon devices, we conduct experiments on three M2 series models. Table~\ref{tab:deviceinfo} shows the size of VRAM/Unified Memory, the single-precision computing power, and the price of each device \cite{applem2pro,applem2ultra,gpuprice}.
We do not include RTX A100 or similar products popular in data centers because of their limited availability, especially when ordered in smaller quantities, and prohibitively high prices.












\begin{table}[t]
	\centering
	\caption{
		Related parameters of each device.
	}
	\includegraphics[width=0.48\textwidth{}]{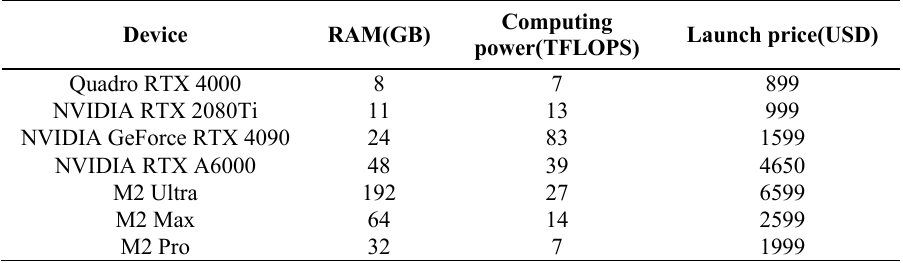}
	\label{tab:deviceinfo}
\end{table}


\subsection{Methodologies}

\begin{figure}[h]
	\includegraphics[width=0.48\textwidth]{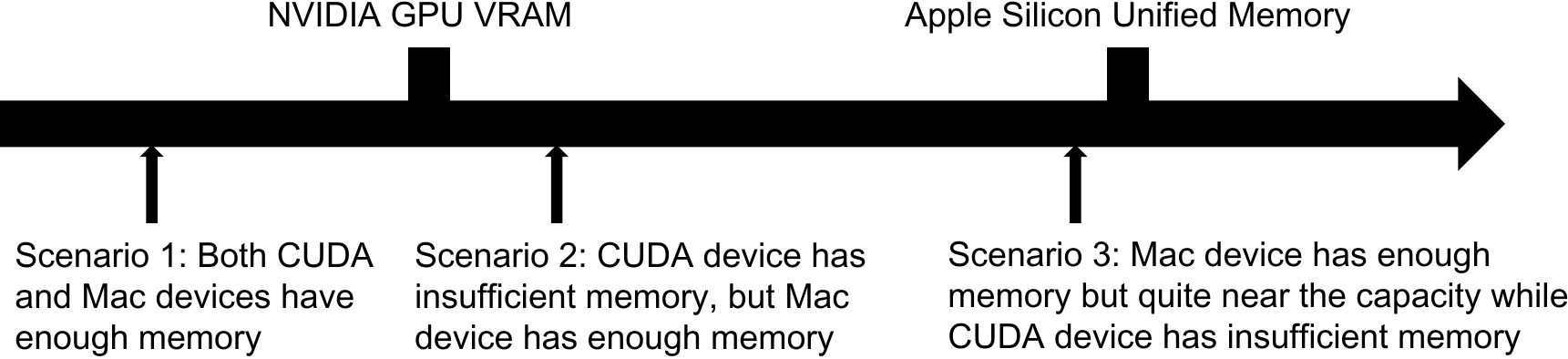}
	\caption{Different scenarios of training. }
	\label{fig:finding:scenariobar}
\end{figure}

\begin{table}[t]
	\centering
	\caption{
		The workloads we used in our experiments, with AdamW optimizer. 
	}
	\includegraphics[width=0.48\textwidth{}]{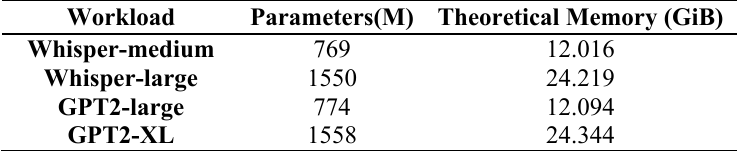}
	\label{tab:workload}
\end{table}

\begin{table}[t]
	\centering
	\caption{
		The configuration of our experiments for end-to-end training on the Hugging Face platform. 
	}
	\includegraphics[width=0.48\textwidth{}]{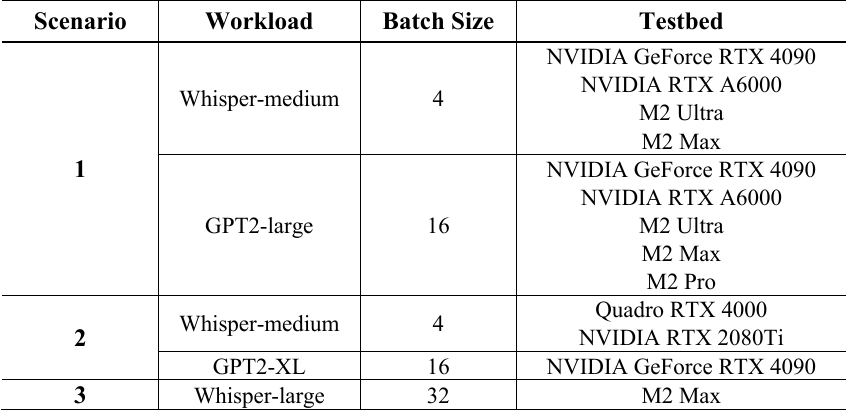}
	\label{tab:e2econfig}
\end{table}
In our experiment, we select a representative set of the most widely used workloads in the training and fine-tuning of large models. Specifically, we conduct experiments on two modalities of data, i.e., speech and text.



For speech, we choose Whisper, an automatic speech recognition model designed to transcribe spoken words into text with high accuracy for multiple languages and dialects~\cite{radford2022robust}. Based on the Transformer architecture~\cite{vaswani2017attention}, Whisper is computationally intensive, requiring a lot of memory and processing resources to manage its large number of parameters and operate efficiently on complex audio data. Due to the model complexity, we pretrain the medium variation of Whisper.

For text, we both choose GPT-2~\cite{radford2019language}, a classic large language model that precedes Whisper but is built on the same Transformer architecture. The computational complexity and memory footprint of GPT-2 is slightly less than that of Whisper, making it a great example of examining whether less computationally capable Apple Silicon devices can be used in pretraining in addition to merely fine-tuning. We experiment on both pretraining and finetuning the large and XL variations of GPT-2.



We evaluated the model pre-training and fine-tuning under different settings. When all of the model and the training data are stored in VRAM or memory, the training process becomes memory-constrained. If the memory is not sufficient, ZeRO-Offload~\cite{ren2021zero}, which can offload part of the model to the CPU or DRAM, etc., needs to be applied to ensure the training can be processed smoothly. We list the total parameters and theoretical memory consumption of each model in \autoref{tab:workload}. In practice, the runtime memory footprint is slightly larger due to the activations and so on.
Depending on the total available memory on the device we evaluate, these workflows fall under different scenarios as introduced in \autoref{fig:finding:scenariobar}. We outline the corresponding memory scenario for each device and workload combination in \autoref{tab:e2econfig}. We adjust the sizes under each memory scenario on all the devices. They are as large as possible for the respective scenario.



We select the time required for a single pass on the same, fixed batch size during training and the memory usage during training as the metrics for our experiments. In particular, for Apple Silicon, we also observe the number of page faults and try to better understand how the unified memory impacts the performance. For energy efficiency, we measure the energy consumption of the Nvidia GPUs, and the GPUs of Apple silicon SoCs.

\section{Findings}

\label{sec:finding}

\subsection{End-to-end evaluations}

\begin{figure*}[h]
	\includegraphics[width=\textwidth]{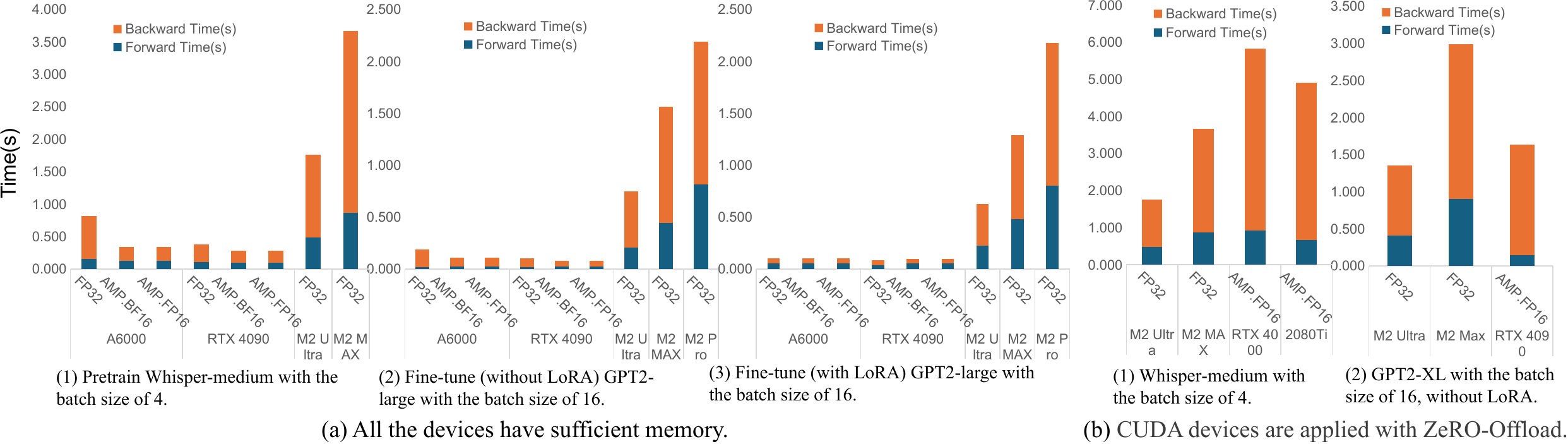}
	\caption{The forward time, backward time on different devices on the two workloads. }
	\label{fig:finding:trainingtime}
\end{figure*}









\paragraph{Precision}
For both CUDA and Mac devices, we select FP32 as training precision. In addition, we try the mixed precision training~\cite{micikevicius2018mixedprecisiontraining}, which is marked as AMP.FP16 and AMP.BF16, respectively, for CUDA only because at this time the torch AMP lacks support for Apple silicon SoCs.

\textbf{Scenario 1: Both CUDA and Apple silicon devices have enough memory} In this scenario, the model and training data can both fit into the Nvidia GPU VRAM and the unified memory of Apple silicon SoC.

In the first benchmark, we pre-train the medium checkpoint of the Whisper model on the Hugging Face platform~\cite{wolf2020huggingfacestransformersstateoftheartnatural}, on CUDA and Mac devices, with the Common Voice dataset~\cite{ardila2020commonvoicemassivelymultilingualspeech}. Figure~\ref{fig:finding:trainingtime}(a)(1) shows the forward and backward time. In the second benchmark, we fine-tune the checkpoint of the GPT2-large model on the Hugging Face platform, on the IMDB dataset~\cite{maas-etal-2011-learning}, with LoRA~\cite{hu2021loralowrankadaptationlarge} and without LoRA respectively. Figure~\ref{fig:finding:trainingtime}(a)(2) and (3) show the forward time and backward time without and with LoRA respectively.  

We have to admit that all Apple silicon SoCs underperform NVIDIA GPUs under this circumstance. The most significant difference between the two kinds of devices during training is memory usage. On all CUDA devices, memory usage is relatively stable, which means that NVIDIA GPUs transfer almost all required data to VRAM at once during training; on Apple Silicon, memory usage(RSS) increases gradually, which means that Apple Silicon gradually transfers required data to Unified Memory during training. 



\begin{table}[t]
	\centering
	\caption{
		The time of each pass during GPT2-large pre-training without the Hugging Face platform.
	}
	\includegraphics[width=0.48\textwidth{}]{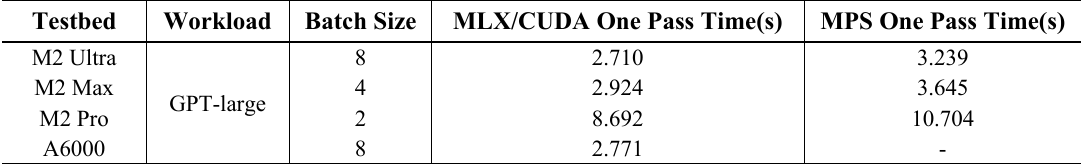}
	\label{tab:mlxe2egpttime}
\end{table}
In addition, we pre-train the GPT2-large with the MLX framework on M2 Ultra, M2 Max, and M2 Pro respectively, using the PTB corpus dataset~\cite{marcus-etal-1993-building}. Unlike the MPS backend, in this setting, the RSS remains stable. For a fair comparison, we furthermore pre-train the GPT2-large with Pytorch MPS or CUDA backend only (i.e. without Hugging Face platform) on M2 Ultra, M2 Max, M2 Pro, and RTX A6000 respectively on the PTB corpus dataset. The time required for each pass with two frameworks is shown in Table~\ref{tab:mlxe2egpttime}, which reveals that the MLX framework has a better performance than the MPS backend on Transformer-based neural network training. Also, in this case, the RTX A6000 slightly underperforms compared to the M2 Ultra(MLX).










\textbf{Scenario 2: CUDA device has insufficient memory but Mac device has enough memory} In this scenario, the model and training data require more memory than the capacity of Nvidia GPU VRAM. When the CUDA device has insufficient memory, in order to ensure that training can proceed normally, we must offload part of the data to the CPU or DRAM, which will bring additional data transmission overhead to the training process. We fine-tune the medium checkpoint of the Whisper model on both CUDA and Mac devices and use the Zero-Offload algorithm to complete the offload operation on CUDA devices to ensure the training process can be completed. The forward and backward time have been reported in Figure~\ref{fig:finding:trainingtime}(b)(1). We also conduct the experiment on the workload of the XL checkpoint of the GPT2 model without LoRA, and the result is also in Figure~\ref{fig:finding:trainingtime}(b)(2).
Through experiments, we can see that the overhead of the data transmission required by ZeRO-Offload on CUDA devices is quite huge. For instance, even on NVIDIA GeForce RTX 4090, if ZeRO-Offload is applied, the training performance is still worse than M2 Ultra.

\begin{figure}[h]
	\includegraphics[width=0.48\textwidth]{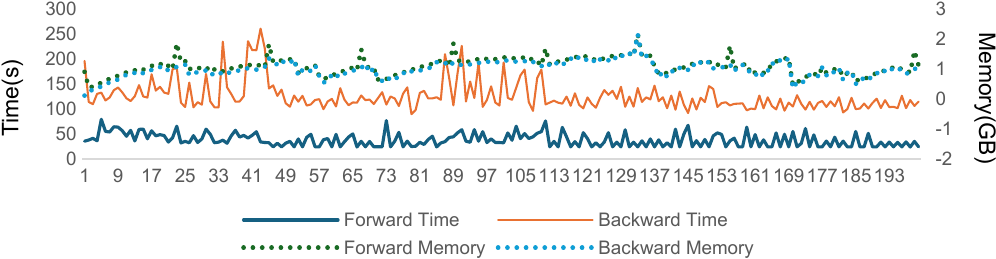}
	\caption{Memory consumption over time during near-capacity training of Whisper-large on M2 Max.}
	\label{fig:finding:nearcapacitymemorywhisperlarge}
\end{figure}

\textbf{Scenario 3: Mac device has enough memory but quite near the capacity} Moreover, we pre-train the large checkpoint of the Whisper model and adjust the batch size to 32 on M2 Max. In this scenario, the Mac device still has enough memory, but the total required memory is very approaching the device capacity. We observe the changes in Figure~\ref{fig:finding:nearcapacitymemorywhisperlarge} during near capacity training and find that RSS is still increasing, but the growth rate is significantly slower than that when memory is super sufficient, and the RSS has maintained at a lower level than the model size. The time required for each pass in Figure~\ref{fig:finding:nearcapacitymemorywhisperlarge} is also observed as a great performance impact and unsteadiness compared to the scenario when the memory is sufficient.



Looking at the above scenarios together, we can draw the following conclusions:
(1) Apple Silicons underperform the other NVIDIA GPUs when the memory footprint required is lower than the capacity of NVIDIA GPUs;
(2) Apple Silicons have a better performance when the VRAM of GPUs is not enough and ZeRO-Offload must be applied.


\subsection{Energy efficiency}
We also measure and record the GPU energy consumption per iteration of each device during the training the medium checkpoint of the Whisper model on the Common Voice dataset with the batch size of 4 and data type of FP32 (in Scenario 1). 


In terms of the energy efficiency of the training, we observe that Apple Silicon demonstrates superior energy efficiency compared to other hardware platforms. As shown in the measurements shown in Table.\ref{tab:power}, the energy consumption per iteration of two Apple Silicon devices is similar to RTX 4090 and much better than A6000. Moreover, considering that the Nvidia GPU platforms need additional power supply for their CPU and memory, which will usually be about 70W~\cite{server-energy}, the gap would be further widened. The improvement in power efficiency not only reduces costs in training but also aligns with sustainable practices by reducing total energy consumption.



\begin{table}[t]
	\centering
	\caption{
		The GPU energy consumption per iteration of each device during the training.
	}
	\includegraphics[width=0.48\textwidth{}]{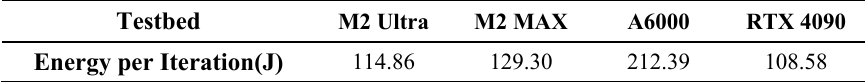}
	
	\label{tab:power}
\end{table}

\subsection{System issues}

To further explain the performance gap in end-to-end training, we will move to the system level.

\paragraph{Memory usage over time}  As mentioned before, Apple Silicons seem to have a greatly different mechanism of memory management from NVIDIA GPUs. Regardless of the workload, Apple Silicon's memory consumption during training shows a gradual increase, rather than maintaining  relatively stable like NVIDIA GPUs, which may be a major reason for the underperformance of Apple Silicon in LLM training.


\paragraph{Page faults over time} To get a better understanding of the overhead of the memory management of Apple Silicon during the training process, we measure the number of page faults of M2 Max when we pre-train the medium checkpoint of the Whisper model in Scenario 1.  With the iteration of passes, the number of page faults is increasing gradually, which indicates that Mac triggers the page fault continuously in an attempt to transfer some data into memory. We also measure the number of page faults of M2 Max during the large checkpoint of the Whisper model pre-training with the batch size of 32 in Scenario 3. In this case, the actual memory required is near the SoC capacity, and more page faults are triggered with a stronger growth trend.






\begin{table}[t]
	\centering
	\caption{
		Kernel launch time on different devices.
	}
	\includegraphics[width=0.48\textwidth{}]{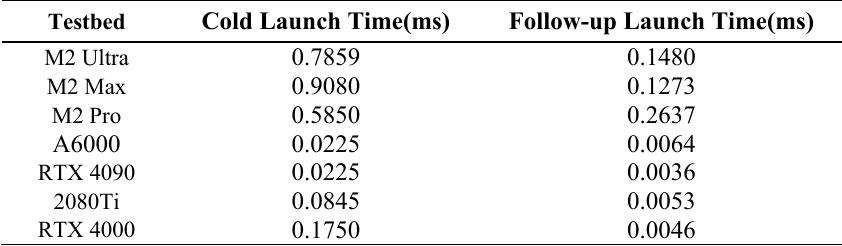}
	\label{tab:kernelLaunchTime}
\end{table}

\paragraph{Overhead of the GPU runtime\label{kernellaunchtime}}
We conduct the experiment to measure the latency of the kernel launch on CUDA devices and Apple devices respectively, which is the layer below the BLAS and MLX. Table~\ref{tab:kernelLaunchTime} shows the kernel launch time during the multiple iterations. The "cold launch" (i.e. the first time launching a kernel)is slower than the following launch and CUDA devices get less latency than Apple Silicon chips. Then when we repeatedly launch the kernel, the latency on CUDA devices is still much lower than that on Apple Silicon. Some common belief may be that Apple Silicon has a more optimized kernel launch time thanks to the unified memory architecture; however, our measurement shows the contract results. Besides, we do not observe any thermal-related issues on Apple silicon devices.





\begin{figure}[h]
	\centering
	\includegraphics[width=0.48\textwidth]{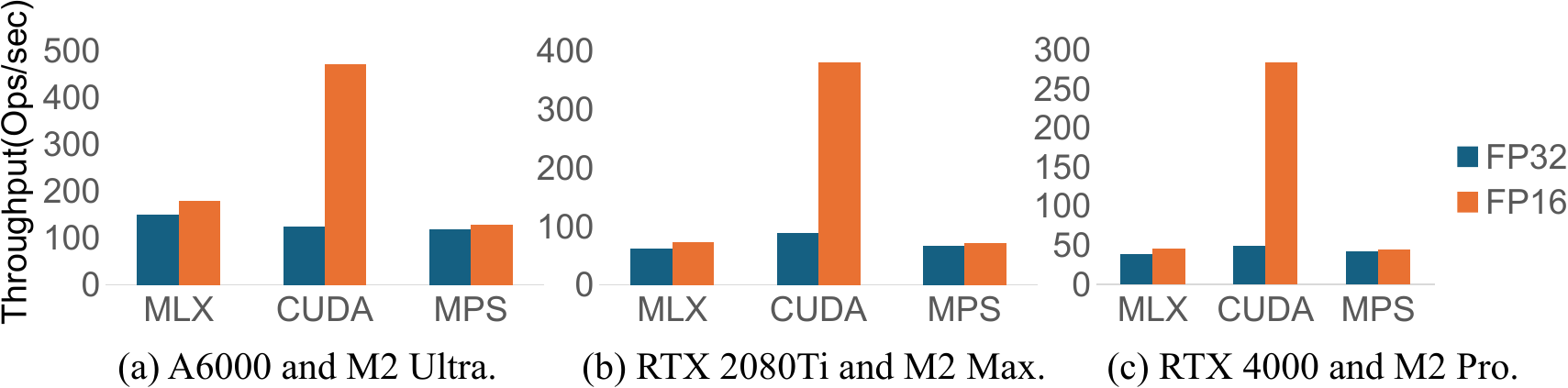}
	\caption{Measurements on Matrix-Matrix product. 
    } 
	\label{fig:finding:mmproduct}
\end{figure}

\begin{figure*}[h]
	\centering
	\includegraphics[width=\textwidth]{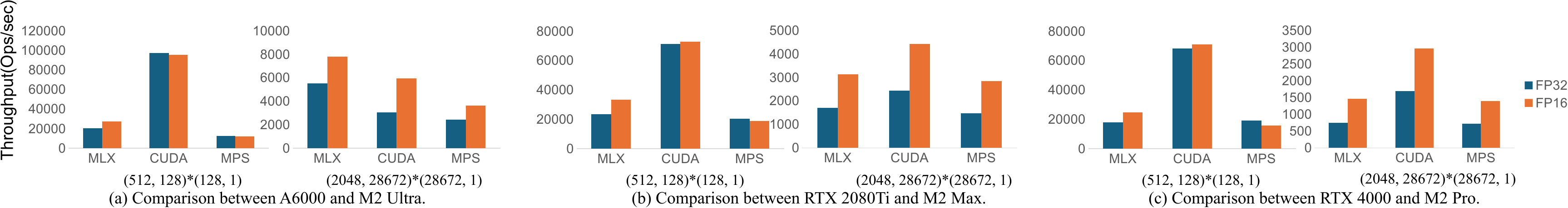}
	\caption{Measurements on Matrix-Vector product.}
	\label{fig:finding:mvproduct}
\end{figure*}

\begin{figure*}[h]
	\centering
	\includegraphics[width=\textwidth]{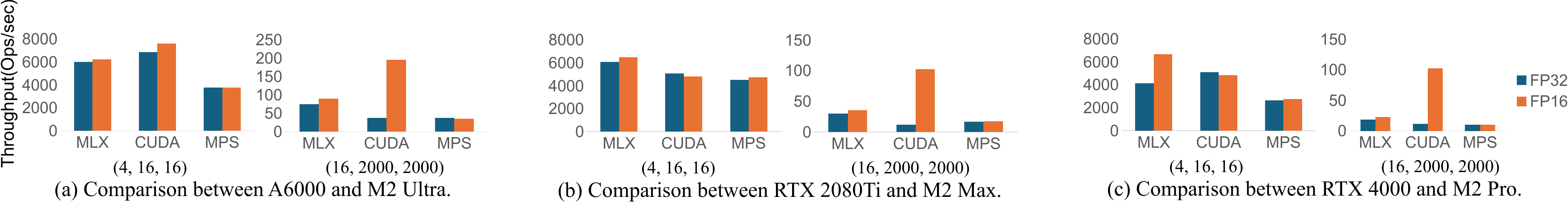}
	\caption{Measurements on Vector-Jacobian Product.}
	\label{fig:finding:vjp}
\end{figure*}

\subsection{BLAS kernel analysis}


To explain the performance gap, we dive into basic linear algebra subprograms (BLAS) performance differences between NVIDIA GPUs and Apple Silicon GPUs. Since the majority of operations of neural networks are on matrices, BLAS performance largely determines the aforementioned performance of end-to-end training~\cite{kepner2017enabling}.

We conduct the experiments on Matrix-Matrix Product, Matrix-Vector Product, and Vector-Jacobian Product respectively across a variety of shapes. Additionally, we benchmark both the FP16 and FP32 data types.
Based on computing power, we compare the computing throughput of A6000 and M2 Ultra, 2080Ti and M2 Max, and RTX4000 and M2 Pro. We use both MPS in Pytorch and MLX framework for Apple Silicon and CUDA in Pytorch for NVIDIA GPUs as the software support.
As we mentioned in \ref{kernellaunchtime}, the cold launch will have an apparently longer latency than the follow-up launch. Thus, in order to eliminate the impact that the cold launch may cause, we first warm up for 10 iterations and then launch the kernel 100 times continuously and repeatedly and report the average of all these results to reduce the effect of unsteadiness.





\paratitle{Matrix-Matrix Product}
We benchmark Batched Matrix-Matrix Product for the batch size of 8 and matrix shapes of (2000*2000), which are common in latest transformer-based LLMs. The benchmark results are shown in Figure \ref{fig:finding:mmproduct}.

NVIDIA GPUs, which have support for Tensor Cores, can get more acceleration when using FP16 instead of FP32, while MPS can hardly get any benefits and MLX can get only about 20\%-30\% benefit.
With FP16, MLX underperforms under all the conditions compared to CUDA but has a better performance than MPS in most cases. This may be because the Tensor Cores of NVIDIA GPUs have very good optimized support for FP16-type operations, and this support is not yet perfect in the Apple Silicon ecosystem. With FP32, the performance gap between MLX or MPS and CUDA is not as large as that with FP16. MLX and MPS shows slightly weaker but largely comparable performances as CUDA.

\paratitle{Matrix-Vector Product}
We also benchmark the Matrix-Vector Product operations with various shapes of matrices and vectors. These operations are also common in transformer-based LLMs, representing some of the small operations when memory bandwidth may be a bottleneck. The benchmark results are shown in Figure~\ref{fig:finding:mvproduct}.

On both CUDA and Apple Silicon devices, using FP16 has a similar ratio of acceleration compared to using FP32 when the shape is relatively large, and using FP16 has similar performance to using FP32 for some small matrices and vectors. For Matrix-Vector Product, Tensor Cores cannot optimize the computation as well as for Matrix-Matrix Product. Additionally, Matrix-Vector Product usually involves smaller data sizes, often hitting memory bandwidth limits rather than computational limits.
Compared with FP16, the performance gap between MLX and CUDA has narrowed, and MLX even outperforms CUDA in some cases. MPS shows a similar performance in the measurement compared to MLX.


\paratitle{Vector-Jacobian Product}
We conduct benchmarking of Vector-Jacobian Product operations, which are the main operations during the backpropagating process, across various shapes of matrices and vectors. The benchmark results are shown in Figure~\ref{fig:finding:vjp}.

In CUDA cases, using FP16 can obtain an obvious(about 5x-6x) acceleration compared to using FP32; but in MLX or MPS cases, using FP16 can reach about 2x acceleration.
With FP32, MLX performs better than CUDA if the shapes are relatively large, but they have similar performance if tensors are in small shapes; with FP16, CUDA performs better than MLX if the shapes are relatively large, but still have similar performance if tensors are in small shapes. However, for the Vector-Jacobian Product kernel, MPS shows an obvious underperformance compared to MLX.



\textbf{Conclusion: }The performance gap observed in BLAS kernels (primary factors), combined with the system issues explored earlier (secondary factors), can explain the performance gap we observed for the end-to-end training. Therefore, we believe we have identified the root causes.


\section{Recommendation}
\label{sec:rcmd}





% ... 







\subsection{For ML practitioners}

If the model is much smaller than a single NVIDIA GPU card (such as 48 GB) then just going with it will be a good choice. If you already have a Mac device with a unified memory larger than the memory footprint of training the model, going ahead and using it will not be too terrible considering that it may cost much more if you buy the NVIDIA GPU cards that can fit the model size, but with an expectation that the speed is about 3x-4x lower than an NVIDIA GPU of similar price. If you want to purchase a new machine dedicated to machine learning training, then Mac devices will be a better choice only if you are about to dive into larger models, but for most general models, NVIDIA GPU will always be a smart choice.
From our observation, thermal is not a big concern. The main concern should be the slow speed when Mac devices compute with FP16 data type.

\subsection{For hardware/software vendors}

Optimized kernels for Apple Silicon achieve only a 30\%–40\% speed advantage over PyTorch, and even after fully porting MLX to PyTorch, a significant performance gap still exists. To make Apple Silicon a more viable and attractive option for training, a road map must address key areas. First, introducing dedicated hardware for FP16 operations, such as Tensor Cores, could significantly enhance performance, though a strong justification for the necessary hardware investments would be required. Next, expanding FP16 support within Automatic Mixed Precision(AMP) would improve efficiency and usability for mixed precision training. Finally, ensuring seamless integration of these advancements into the Apple Silicon training ecosystem would be essential to realize their full potential.

\subsection{For ML benchmark development}

For future tests on Apple Silicon, the training performance with FP16 should be checked; if the software support for FP16 remains lacking, the performance gap will likely persist.

\section{Concluding remarks}

This paper has measured the performance of LLM training on the Apple Silicon and made comparisons side-by-side with some NVIDIA GPUs with similar prices. We also try to explain the performance gap from the system level and dive into BLAS kernels to figure out where the gap is from. As a result, based on our observation, we have some recommendations from various perspectives.


\newpage




\bibliographystyle{plain}
\bibliography{bib/abr-short,bib/xzl,bib/hongyu,bib/misc,bib/book,bib/security,bib/iot,bib/datacentric,bib/hp,bib/transkernel,bib/ml-edge,bib/secureGPU,bib/TrustZone,bib/lwg,bib/sysml,bib/mm,bib/xzm}



\end{document}